\documentclass[letterpaper, 10 pt, conference]{ieeeconf}  
\pdfoutput=1

\IEEEoverridecommandlockouts                              

\usepackage{graphics, subfig} 
\usepackage{epsfig} 
\usepackage{mathptmx} 
\usepackage{times} 
\usepackage{amsmath} 
\usepackage{amssymb}  
\usepackage{algorithmic,algorithm}
\usepackage{comment}
\usepackage{breqn}
\usepackage{xcolor}

{\renewcommand{\baselinestretch}{1}}

\usepackage{wrapfig}

\begin{document}

\title{\LARGE \bf
Towards Efficient Modularity in Industrial Drying: A Combinatorial Optimization Viewpoint
}
\author{Alisina Bayati$^{1a}$, Amber Srivastava$^{2}$, Amir Malvandi$^{3}$, Hao Feng$^{4}$ and Srinivasa M. Salapaka$^{1b}$
\thanks{$^{1}$Department of Mechanical Science and Engineering, University of Illinois at Urbana-Champaign, 61801 IL, USA. $^{a}${\tt\small abayati2@illinois.edu}, $^{b}${\tt\small salapaka@illinois.edu}}
\thanks{$^{2}$Automatic Control Laboratory, Swiss Federal Institute of Technology (ETH Zurich), Physicstrasse 3, 8092 Zurich, Switzerland. {\tt\small asrivastava@ethz.ch}}
\thanks{$^{3}$Department of Agriculture and Biological Engineering Sciences, University of Illinois at Urbana-Champaign, 61801 IL, USA. $^{3}${\tt\small amirm2@illinois.edu}}
\thanks{$^{3}$North Carolina Agricultural and Technical State University, 27411 NC. {\tt\small hfeng@ncat.edu}}
\thanks{This work was supported by the U.S. Department of Energy under award DE-EE0009125 and NCCR Automation (grant number 180545) funded by the Swiss National Science Foundation.}}

\thispagestyle{empty}
\pagestyle{empty}

\maketitle

\begin{abstract}
The industrial drying process consumes approximately 12\% of the total energy used in manufacturing, with the potential for a 40\% reduction in energy usage through improved process controls and the development of new drying technologies. To achieve cost-efficient and high-performing drying, multiple drying technologies can be combined in a modular fashion with optimal sequencing and control parameters for each. This paper presents a mathematical formulation of this optimization problem and proposes a framework based on the Maximum Entropy Principle (MEP) to simultaneously solve for both optimal values of control parameters and optimal sequence. The proposed algorithm addresses the combinatorial optimization problem with a non-convex cost function riddled with multiple poor local minima. Simulation results on drying distillers dried grain (DDG) products show up to 12\% improvement in energy consumption compared to the most efficient single-stage drying process. The proposed algorithm converges to local minima and is designed heuristically to reach the global minimum.
\end{abstract}

\section{Introduction}\label{sec:Intro}
Industrial drying is responsible for roughly 12\% of the total end-use energy used in manufacturing, equivalent to 1.2 quads annually \cite{DOEBarrier}. The US Department of Energy estimates that by implementing more efficient process controls and new drying technologies, it is possible to reduce this amount by approximately 40\% (0.5 quads/year), resulting in operating cost savings of up to \$8 billion per year \cite{DOEReview}. Moreover, the drying process has a significant impact on the quality of food products. Prolonged exposure to excessive heat can have negative effects on the physical and nutritional properties of the products \cite{NutritionalP}.

In recent years, several more efficient drying technologies have been proposed in the literature, such as Dielectrophoresis (DEP) \cite{DEP2}, ultrasound drying (US) \cite{US}\cite{US2}, slot jet reattachment nozzle (SJR) \cite{SJR}, and infrared (IR) drying \cite{IR}. These technologies have helped improve product quality and energy efficiency. Industrial drying units typically use one of these technologies to achieve their drying goals. However, each technology performs with different efficiencies in different settings. Depending on the operating conditions, some technologies may be more favorable than others. For example, contact-based ultrasound technology is more effective in the initial phase of the process, where the moisture content of the food sample is relatively high, while pure hot air drying consumes less energy and is more effective when the moisture content is low. By combining these two processes, it is possible to take advantage of both technologies and compensate for their inefficiencies. Therefore, understanding (a) the sequence in which different drying techniques should be used, and (b) the operating parameters of each technology, can help us maximize their capabilities. Dividing the drying process into sub-processes that use different drying methods and operating conditions can help alleviate their individual limitations.

\captionsetup[figure]{font=small,labelfont=small}
\begin{figure}[t]
\centering
  \includegraphics[width=0.5\textwidth]{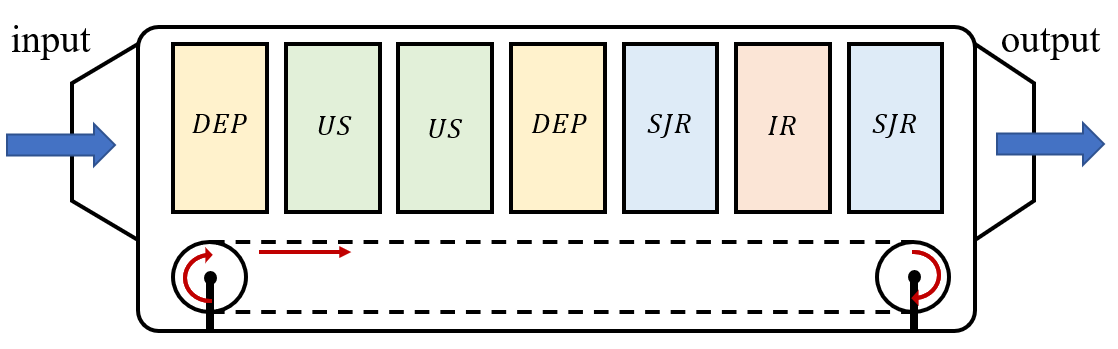}
  \caption{Schematics of the continuous \textit{smart} dryer prototype with seven buckets which accommodates multiple drying  technologies to achieve better performance. In the example shown above, two DEP, two ultrasound, one IR, and two SJR modules are used in a specific order.}
   \label{fig:SmartD}
\end{figure}
 
To illustrate, let us consider the continuous drying testbed depicted in Fig.\ref{fig:SmartD}, which includes several drying modules such as ultrasound, DEP, SJR nozzle, and IR technologies. Each drying module is controlled by a set of parameters that influence the amount of moisture removal. For instance, ultrasound power and duty cycle are the control parameters of the ultrasound technology, while electric field intensity is the control variable of the DEP module. Additionally, the control parameters of each dryer impact the amount of energy consumed during the process, creating a tradeoff between energy consumption and moisture removal. Therefore, to minimize the total energy consumed by the testbed while achieving the desired moisture removal, a combinatorial optimization problem can be formulated to determine the optimal order in which the drying modules should be placed in the testbed and the optimal control parameters associated with them.

\begin{figure}[t]
\centering
  \includegraphics[width=0.5\textwidth]{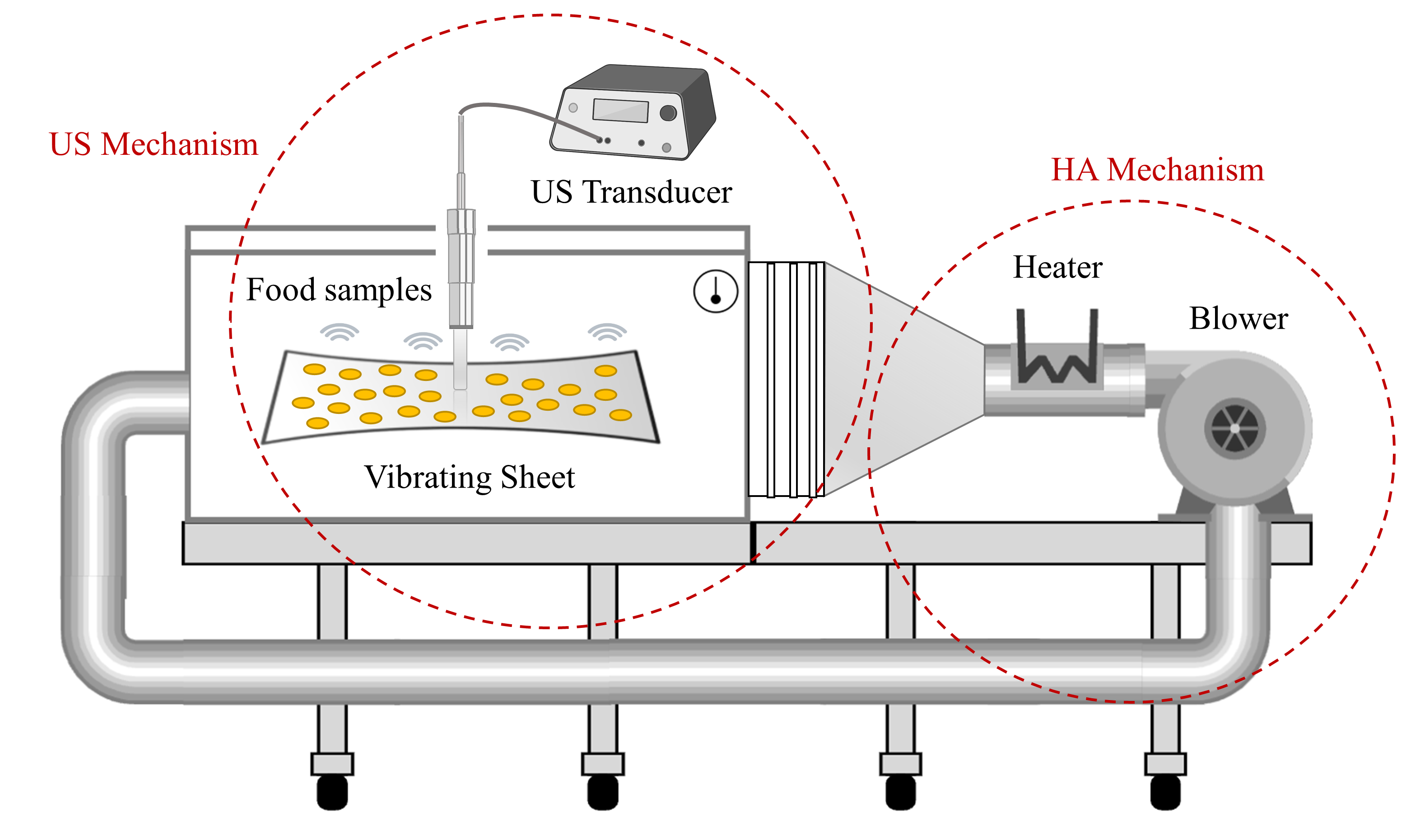}
  \caption{Schematics of the convective/ultrasound testbed for batch-process drying which can be used for both pure hot air (HA) and combined hot air and ultrasound (HA/US) processes. The HA mechanism consists of a blower and a heater, whereas the US mechanism is a vibrating sheet attached to the ultrasound transducer. One can switch from HA/US process to pure HA process by turning off the US transducer}
   \label{fig:Batch_p}
\end{figure}

Similarly, this approach can be extended to batch-process drying with some adjustments. For example, in the testbed shown in Fig.\ref{fig:Batch_p}, which is used for the batch drying process, each technology can be used more than once. It includes an ultrasonic module \cite{US}, a drying chamber with a rectangular cross-section, a blower, and a heater. The food sample is located on a vibrating sheet attached to the ultrasonic transducer and exposed to the hot air coming from the heater, allowing combined hot-air and ultrasound drying. In this setup, the problem of interest is to reduce energy consumption, if possible, by dividing the process into consecutive pure hot-air (HA) and combined hot-air and ultrasound (HA/US) sub-processes, each with different operating conditions.

\captionsetup[figure]{font=small,labelfont=small}

Previous research in the field of drying has largely focused on improving the efficiency of existing drying methods or developing new technologies \cite{DEP,US,SJR}. Some studies have used optimization routines such as the response surface method (RSM), a statistical procedure, to optimize process control variables using experimental data \cite{RSM2,RSM3}. However, there is limited literature that addresses optimization problems related to integrating different drying technologies through sequencing and parameter optimization. The primary contribution of our work is the modular use of multiple existing technologies to achieve cost efficiency with desired performance levels, while also allowing for optimal operating conditions that can vary over time, potentially improving performance even further. In our simulation results, presented in Section \ref{sec:SR}, we show up to a 12\% reduction in energy consumption compared to the most efficient single-stage hot-air/ultrasound drying process, as well as up to a 63\% improvement in energy efficiency compared to the commonly used optimal hot air drying method. Similar optimization problems can arise in various industrial processes that involve using a sequence of distinct devices with similar functions to form a unified process, such as the wood pulp industry with drying drums varying in radius and temperature, route optimization in multi-channel wireless networks with heterogeneous routers, and sensor network placement.
 
This paper introduces a framework based on the Maximum Entropy Principle (MEP) to model and optimize the various sub-processes in an industrial drying unit. These optimization problems pose significant challenges due to the combinatorially large number of valid sequences of sub-processes and their discrete nature. To address these issues, we assign a probability distribution to the space of all possible configurations. However, determining the optimal operating conditions of sub-processes alone is analogous to the NP-hard resource allocation problem, with a non-convex cost surface containing multiple poor local minima. Traditional algorithms like k-means often get trapped in these local minima and are sensitive to initialization. To overcome this, our algorithm uses a homotopy approach from an auxiliary function to the original non-convex cost function. This auxiliary function is a weighted linear combination of the original non-convex cost function and an appropriate convex function, chosen as the negative Shannon entropy of the probability distribution defined above. We start with weights that favor the negative Shannon entropy term, making the function convex and easily solvable. As the iteration progresses, the weight of the original non-convex cost increases, and the obtained local minima are used to initialize subsequent iterations. The auxiliary function converges to the original non-convex cost function at the end of the procedure. This approach is independent of initialization and tracks the global minimum by gradually transforming the convex cost function to the desired non-convex cost function.

\section{Problem Formulation}\label{sec:PF}

\begin{figure}[t]
\centering
  \includegraphics[width=0.5\textwidth]{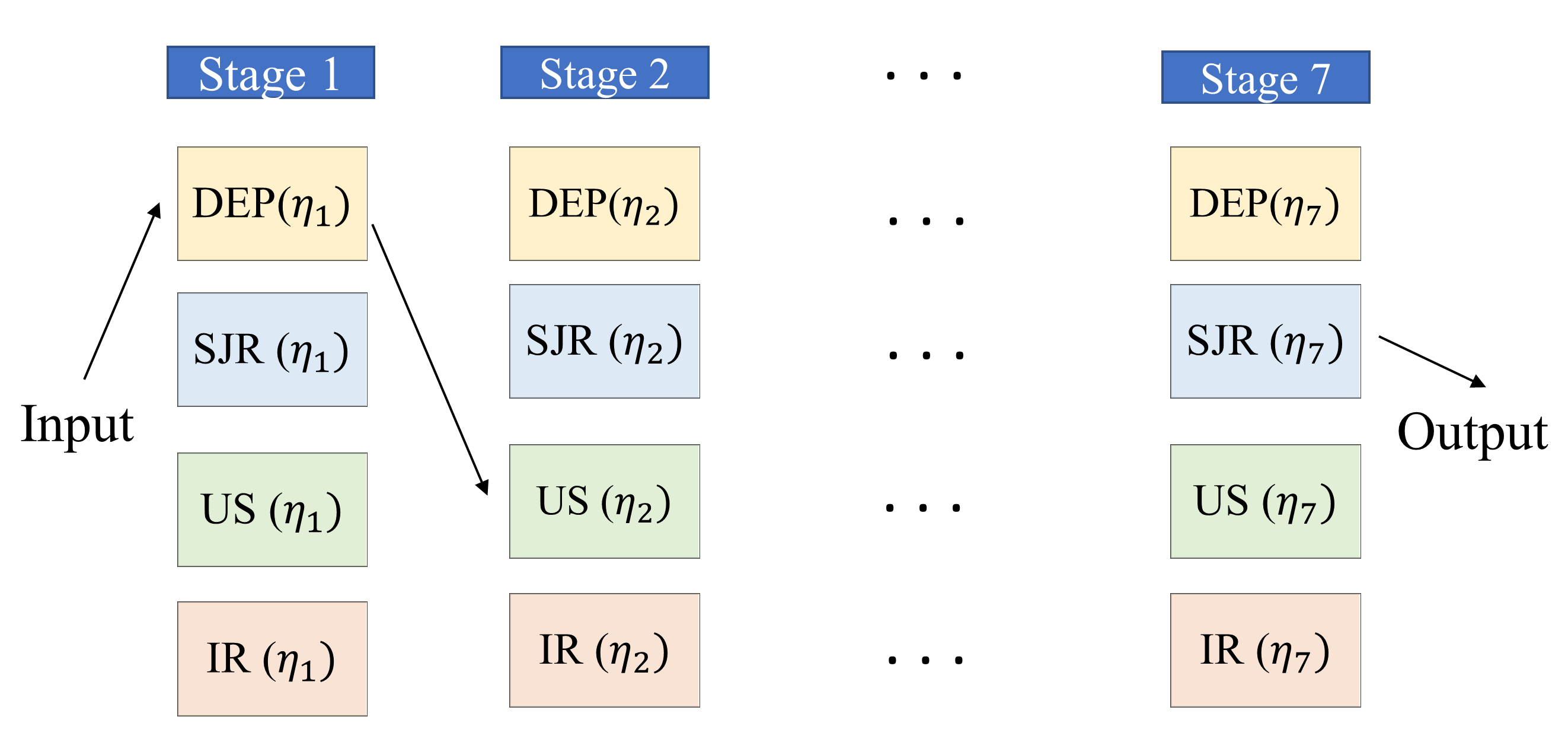}
  \caption{Diagram corresponding to the process shown in Fig. \ref{fig:SmartD} in which $\Gamma = \{\text{DEP (1), SJR (2), US (3), IR (4)}\}$. In the sequence shown by the arrows, $\gamma_1 =1, \gamma_2 = 3, ... , \gamma_7 = 2$ which defines the process configuration $\omega = (\gamma_1,\gamma_2,...,\gamma_7)$. Moreover, $\eta_i (1 \leq i \leq7)$ determines the control variables of the technology used in $i-$th stage.}
   \label{fig:Seq_Dec}
\end{figure}

We formulate the problem stated above as a \textit{parameterized path-based optimization problem} \cite{Kale}. Such problems are described by a tuple
\begin{align}
\mathcal{M}=\langle M,\gamma_1,\hdots,\gamma_M,\eta_1,\hdots,\eta_M,D\rangle,
\end{align}
where $M$ is the number of stages allowed, and $\gamma_k$ denotes the sub-process chosen to be used in the $k-$th stage. In particular,
\begin{align}
\gamma_k \in \Gamma_k := \{f_{k1},\hdots,f_{kL_k}\} \quad  \forall 1 \leq k \leq M,
\end{align}
where $\Gamma_k$ is the set of all sub-processes permissible in the $k-$th stage. Moreover,
\begin{align}
&\eta_k \in H(\gamma_k) \subseteq \mathbb{R}^{d_{\gamma_k}} \quad  \forall  1 \leq k \leq M,
\end{align}
where $\eta_k$ and $H(\gamma_k)$ denote the control parameters associated with the $k-$th sub-process and its feasible set, respectively.
$D(\omega,\eta_1,\hdots,\eta_M)$ denotes the cost incurred along a {\em path} $\omega$, where $\omega\in\Omega := \{(f_{1i_1},f_{2i_2},\hdots,f_{Mi_M}):f_{ki_k}\in\Gamma_k\}$ represents a sequence of sub-processes starting from the first stage to the terminal stage $M$. The objective of the underlying parameterized path-based optimization problem is to determine (a) the optimal path $\omega^*\in\Omega$, and (b) the parameters $\eta_k^*$ for all $1\leq k \leq M$ that solves the following optimization problem 
\begin{equation}
\begin{aligned}
\min_{\{\eta_k\},\nu(\omega)} \label{eq: path-based}
&\sum_{\omega\in\Omega}\nu(\omega)D(\omega,\eta_1,\hdots,\eta_M),\\
\text{subject to }& \sum_{\omega\in\Omega} \nu(\omega) = 1,~~ \nu(\omega)\in\{0,1\}\\
& \eta_k \in H(\gamma_k) \quad \forall 1 \leq k \leq M,
\end{aligned}
\end{equation}
where $\nu(\omega)$ determines whether or not the path $\omega$ has been taken. In other words,
\begin{align}
\begin{split} \label{eq:nu}
\nu(\omega) = \begin{cases}
1 \quad \text{if $\omega$ is chosen}\\
0 \quad \text{otherwise.}
\end{cases}
\centering
\end{split}
\end{align}\\
Fig. \ref{fig:Seq_Dec} further illustrates all the notations defined, for the exemplary process shown in Fig. \ref{fig:SmartD}.

One approach to address the optimization problem stated in (\ref{eq: path-based}) is to solve each objective separately. However, in this approach, the coupledness of the two objectives is not taken into account which may result in a sub-optimal solution. On the other hand, our MEP-based approach aims for solving the two simultaneously.

Let us reconsider the batch-process drying example described earlier, where the testbed allows up to $M$ different sub-processes, each could be either HA or HA/US. To pose the problem of interest as a parameterized path-based optimization problem, we define
\begin{align}
\Gamma_k := \{0,1\} \quad \forall 1 \leq k \leq M,
\end{align}
in which 0 and 1 indicate HA and HA/US sub-processes, respectively. Thus, the process configuration $\omega \in \Omega$ would become
\begin{equation} \label{eq:gamma}
\begin{aligned}
\omega = (\gamma_1, \gamma_2, ... , \gamma_M), \quad \gamma_k \in \{0,1\} \quad \forall 1 \leq k \leq M,
\end{aligned}
\end{equation}

The control parameters of both sub-processes, in this case, are residence time $t$ and air temperature $T$. The heater of the setup in Fig. \ref{fig:Batch_p} is designed to keep the air temperature between $30^{\circ}C$ and $70^{\circ}C$. Also, considering the settling time of the air temperature, it is required for all sub-processes to take at least $t_0 = 2$ minutes. Hence,
\begin{align}
&\eta_k = \begin{bmatrix} t_k\\T_k
 \end{bmatrix} \in U \quad \forall 1 \leq k \leq M,
\end{align}
where $U$ is defined as below and denotes the set of all admissible control parameters.
\begin{align}
 U := \left \{ \begin{bmatrix} t\\T
 \end{bmatrix} \in \mathbb{R}^2: t \geq 2 \; \text{mins} \;, T \in [30, 70]^{\circ} C
 \right\}\label{eq:constraint_set}
\end{align}
A key assumption here is that all the samples within a batch are similar in properties 
 such as porosity and initial moisture content.

\captionsetup[figure]{font=small,labelfont=small}

To determine the cost of a process, we must identify the desired properties of dried food products, such as wet basis moisture content and color. For simplicity, we focus on ensuring that the wet basis moisture content falls within a predetermined range ($\leq x_d$) by the end of the process. In this paper, we denote the wet basis moisture content of the food sample at the end of the k-th stage under process configuration $\omega$ as $x^{(\omega)}_k$. To account for the cost associated with the final moisture content, the corresponding dynamics must be modeled.
\begin{equation} \label{eq:Dynamics}
\begin{aligned}
x_{k}^{(\omega)} = f_{\gamma_k}(x_{k-1}^{(\omega)},\eta_k) , \quad \forall \; 1 \leq k \leq M,
\end{aligned}
\end{equation}
where $x_0^{(\omega)}$ denotes the initial wet basis moisture content of the food sample. The semi-empirical drying curves (moisture content versus time) of distillers dried grains (DDG) were derived and evaluated in \cite{US} for $T=25^{\circ}C$, $T=50^{\circ}C$, and $T=70^{\circ}C$. The kinetics of drying for other temperatures can be approximated by interpolating the experimental drying curves in \cite{US}: 
\begin{equation*} \label{eq:Stat_M}
\begin{aligned}
x_{k+1}^{(\omega)} = \frac{e^{-K_{\gamma_k}(T_k)t_k^*(T_k,x_k^{(\omega)},t_k)} (3.16 - M_{\gamma_k}(T_k)) + M_{\gamma_k}(T_k)}{1 + e^{-K_{\gamma_k}(T_k)t_k^*(T_k,x_k^{(\omega)},t_k)} (3.16 - M_{\gamma_k}(T_k)) + M_{\gamma_k}(T_k)}
\end{aligned}
\end{equation*}
in which\\
\begin{equation} \label{eq:K}
\begin{aligned}
K_0(T_k) = &(0.074493T_k^2 - 45.5058T_k + 6839.9)/1000\\
K_1(T_k) = &(-0.05811T_k^2 + 39.962T_k - 6680.1)/1000\\
\end{aligned}
\end{equation}
($T_k$ in Kelvin) denote the Lewis model constants \cite{Lewis} for HA and HA/US sub-processes, respectively. Also,\\
\begin{equation} \label{eq:Meq}
\begin{aligned}
M_0(T_k) = &(0.2479T_k^2 - 172.09T_k + 30133)/10000\\
M_1(T_k) = &(0.1468T_k^2 - 107.27T_k + 19720)/10000\\
\end{aligned}
\end{equation}
($T_k$ in Kelvin) represent the equilibrium dry basis moisture content (ratio of the weight of water to the weight of the solid material) of the DDG products. Moreover, $t_k^*(T_k,x_k,t_k)$ is defined to be:
\begin{equation} \label{eq:Meq}
\begin{aligned}
t_k^*(T_k,x_k^{(\omega)},t_k) = \frac{1}{K_{\gamma_k}(T_k)}\log(\frac{3.16-M_{\gamma_k}(T_k)}{\frac{x_k^{(\omega)}}{1 - x_k^{(\omega)}} - M_{\gamma_k}(T_k)}) + t_k
\end{aligned}
\end{equation}
Therefore, we define the process cost $D(\omega,\eta_1,\hdots,\eta_M)$ as:
\begin{equation} \label{eq:Cost}
\begin{aligned}
D(\omega, \eta_1, ..., \eta_M) = \sum_{k=1}^M g_{\gamma_k}(\eta_k) + G(x_M^{(\omega)},x_d) \hspace{0.1cm},
\end{aligned}
\end{equation}
in which $g_{\gamma_k}:\mathbb{R}^{d_{\gamma_k}} \rightarrow \mathbb{R}$ is the cost (e.g. energy consumption) of the $k-$th sub-process, and $G: \mathbb{R}^2 \rightarrow \mathbb{R}$ is a function penalizing the violation of the constraint. 
In batch process drying example, $g_0(.)$ is the energy consumed for HA drying which can be approximated using the following:
\begin{equation} \label{eq:Cost_HA}
\begin{aligned}
g_0(\eta_k) = g_0(t_k,T_k) \propto & \hspace{0.1cm} \dot{m}_{air} c_p (T_k - T_0) t_k
\end{aligned}
\end{equation}
where $\dot{m}_{air}$ is the mass flow rate of the inlet air, $T_0$ is ambient air temperature, $A$ is the cross-section area of the chamber, $V_{air}$ is air velocity, $c_p$ and $\rho$ are  average specific heat capacity and density of air in the temperature operating range of the testbed. Therefore, we can use a weighting coefficient $\alpha$ to adjust the cost of the HA sub-process.
\begin{equation} \label{eq:Cost_HA}
\begin{aligned}
g_0(t_k,T_k) = \hspace{0.1cm} \alpha \rho_{air} A V_{air} c_p (T_k - T_0) t_k \hspace{0.1cm},
\end{aligned}
\end{equation}
On the other hand, $g_1(.)$ is the energy consumed for HA/US sub-processes and can be computed using
\begin{equation} \label{eq:Cost_HAUS}
\begin{aligned}
g_1(\eta_k) = g_1(t_k,T_k) = g_0(t_k,T_k) + P_{US}t_k \hspace{0.1cm},
\end{aligned}
\end{equation}
in which $P_{US}$ is the power consumption of the ultrasound transducer. Therefore, the total cost of the process can be written in the following way:
\begin{equation} \label{eq:TotalCost}
\begin{aligned}
&D(\omega, \eta_1, ..., \eta_M) = \\
&\sum_{k=1}^M (\alpha \dot{m}_{air} c_p (T_k - T_0) + \gamma_k P_{US})t_k + G(x_M^{(\omega)},x_d)
\end{aligned}
\end{equation}
As a result, we write the corresponding combinatorial optimization problem below:
\begin{equation}
   \begin{aligned} \label{eq:constraints1}
    \min_{\omega,\{\eta_k\}} &~ D(\omega, \eta_1, \eta_2, ..., \eta_M)\\
     \text{subject to: } \; &\eta_k \in U \quad \forall 1\leq k \leq M
    \end{aligned} 
\end{equation}
To adapt the above problem to the form of the parameterized path-based optimization problem described in (\ref{eq: path-based}), we rewrite it as below:
\begin{equation}
\begin{aligned}
\min_{\{\eta_k\},\nu(\omega)} \label{eq:Optimization2}
&\sum_{\omega\in\Omega}\nu(\omega)D(\omega,\eta_1,\hdots,\eta_M),\\
\text{subject to: }& \sum_{\omega\in\Omega} \nu(\omega) = 1,~~ \nu(\omega)\in\{0,1\}\\
& \eta_k \in U \quad \forall 1 \leq k \leq M,
\end{aligned}
\end{equation}

\section{Problem Solution}\label{sec:PS}

Combinatorial optimization techniques can be used to solve the optimization problem stated in (\ref{eq:Optimization2}). Intuitively, we can view it as a clustering problem in which the goal is to assign a particular sequence of sub-processes to every food sample with known initial moisture content. In this case, the location of the cluster centers can be thought of as the control parameters associated with that sequence. This work, similar to \cite{Kale} and \cite{9096570}, utilizes the idea of the Maximum Entropy Principle (MEP) \cite{jaynes1957information}\cite{jaynes2003probability}. To be able to invoke MEP, we relax the constraint $v(\omega) \in \{0,1\}$ in (\ref{eq:Optimization2}) and let it take any value in $[0,1]$. We denote this new weighting parameter by
\begin{align}
p(\omega) \in [0,1] \quad \forall \omega \in \Omega
\end{align}
In other words, we allow partial assignment of process configurations to the food sample. Note that this relaxation is only used in the intermediate stages of our proposed approach. The final solution still satisfies $p(\omega) \in \{0,1\}$. Without loss of generality, we assume that $\sum_{\omega \in \Omega}p(\omega) = 1$. Hence, we can rewrite (\ref{eq:Optimization2}) as:
\begin{equation}
\begin{aligned}
\min_{\{\eta_k\},p(\omega)} \label{eq:Optimization_relaxed}
&\sum_{\omega\in\Omega}p(\omega)D(\omega,\eta_1,\hdots,\eta_M),\\
\text{subject to: }& \sum_{\omega\in\Omega} p(\omega) = 1,~~ p(\omega)\in [0,1],\\
& \eta_k \in U \quad \forall 1 \leq k \leq M
\end{aligned}
\end{equation}

Since the framework we are presenting is built upon MEP, let us briefly review it in the context of this problem. MEP states that given prior information about the process, the most unbiased set of weights is the one that has the maximum Shannon entropy. Assume the information we have about the process is the expected value of the process cost ($\mathbb{E}(D) = D_0$). Then, according to MEP, the most unbiased weighting parameters solve the optimization problem
\begin{align}  \label{eq:MEP}
\begin{split}
\max_{p} &~  -\sum_{\omega \in \Omega} p(\omega) \log \left ( p(\omega)\right )  \\
\text{subject to:} &~ \Bar{D} = D_0
\end{split}
\end{align}
where $\Bar{D}$ is the expected value of the cost $D$, namely,
\begin{dmath*} \label{Expected_D}
 \Bar{D} = \sum_{\omega \in \Omega} p(\omega) D(\omega, \eta_1, \hdots , \eta_M)   
\end{dmath*} \label{Expected_D}
The Lagrangian corresponding to (\ref{eq:MEP}) is given by the maximization of $H - \beta \Bar{D}$, or equivalently, minimization of $F = \Bar{D} - \frac{1}{\beta}H$, where $\beta$ is the Lagrange multiplier. Therefore, the problem reduces to minimizing $F$ with respect to $p(\omega)$ and $\{\eta_k\}$ such that $\sum_{\omega \in \Omega} p(\omega) = 1$. We add the last constraint with the corresponding Lagrange multiplier $\mu$ to the objective function $F$ and rewrite the problem as below:
\begin{align}
\begin{split} \label{eq:Unc_Lagrangian}
\min_{\{\eta_k\},p(\omega)} &~ \Bar{D} - \frac{1}{\beta}H + \mu (\sum_{\omega \in \Omega}p(\omega) - 1)\\
\text{subject to: }~ &\eta_k \in U \quad \forall 1 \leq k \leq  M
\end{split}
\end{align}
We denote the new objective function in (\ref{eq:Unc_Lagrangian}) by $\Bar{F}$. Note that $\Bar{F}$ is convex in $p$ and therefore, the optimal weights can be determined by setting $\frac{\partial \Bar{F}}{\partial p}=0$ which gives the Gibbs distribution
\begin{align}
\begin{split} \label{eq:p_opt}
p^*(\omega) = \frac{\exp \left( -\beta D(\omega, \eta_1, \hdots, \eta_M)\right)}{\sum_{\omega^\prime \in \Omega} \exp \left( -\beta D(\omega\prime, \eta_1, \hdots, \eta_M)\right)}
\end{split}
\end{align}
Therefore, by  plugging (\ref{eq:p_opt}) into $\Bar{F}$, we obtain its corresponding minimum $\Bar{F}^*$.
\begin{align}
\begin{split} \label{eq:F_star}
 \Bar{F}^* &~= \min_{p(\omega)} \Bar{F}= -\frac{1}{\beta} \log \sum_{\omega \in \Omega} \exp \left( -\beta D(\omega, \eta_1, \hdots, \eta_M)\right)
\end{split}
\end{align}
Subsequently, to determine the optimal process parameters, we minimize $\Bar{F}^*$ with respect to $\eta_k$s. In other words, solving the constrained optimization problem
\begin{align}
\begin{split} \label{eq:eta_opt}
\min_{\{\eta_k\}} &~ \Bar{F}^*\\
\text{subject to: }~ & \eta_k \in U\quad \forall 1 \leq k \leq M 
\end{split}
\end{align}
results in finding the optimal control parameters $\{\eta_k^*\}$ for all the sub-processes. Any constrained optimization algorithm can be used to solve (\ref{eq:eta_opt}). As an example, we used the interior point algorithm in our simulations.\\
The proposed algorithm, thus, consists of iterations with the following two steps:
\begin{enumerate}
    \item Using parameters $\{\eta_k\}$ obtained in the previous iteration to find the optimal weights according to (\ref{eq:p_opt}).
    \item Solving the constrained optimization in (\ref{eq:eta_opt}) to find the optimal parameters $\{\eta_k^*\}$ using $\{\eta_k\}$ as the initial guess for the algorithm.
\end{enumerate}

\captionsetup[figure]{font=small,labelfont=small}
\begin{figure*}[t]
\centering
  \includegraphics[width=1\textwidth]{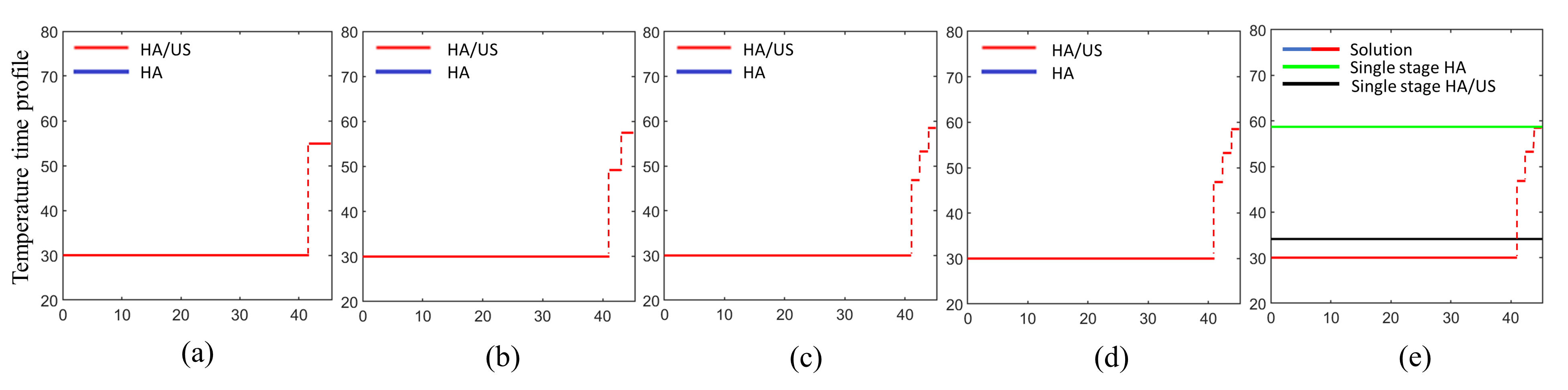}
  \caption{Figures (a), (b), (c), and (d) represent the solution obtained using our proposed algorithm for $M=2$, $M=3$, $M=4$, and $M=5$ accordingly, where $M$ indicates the maximum number of stages allowed. In all simulations, $\alpha$ is chosen to be 0.5. Figure (e) compares the temperature profile of the solution for $M=6$ and optimal single-stage HA and single-stage HA/US processes. The results show a 12.09\% energy consumption reduction compared to the single-stage HA/US process and a 63.19\% improvement compared to the single-stage pure HA process. }
   \label{fig:gamma0.5}
\end{figure*}

\captionsetup[figure]{font=small,labelfont=small}
\begin{figure*}[h]
\centering
  \includegraphics[width=1\textwidth]{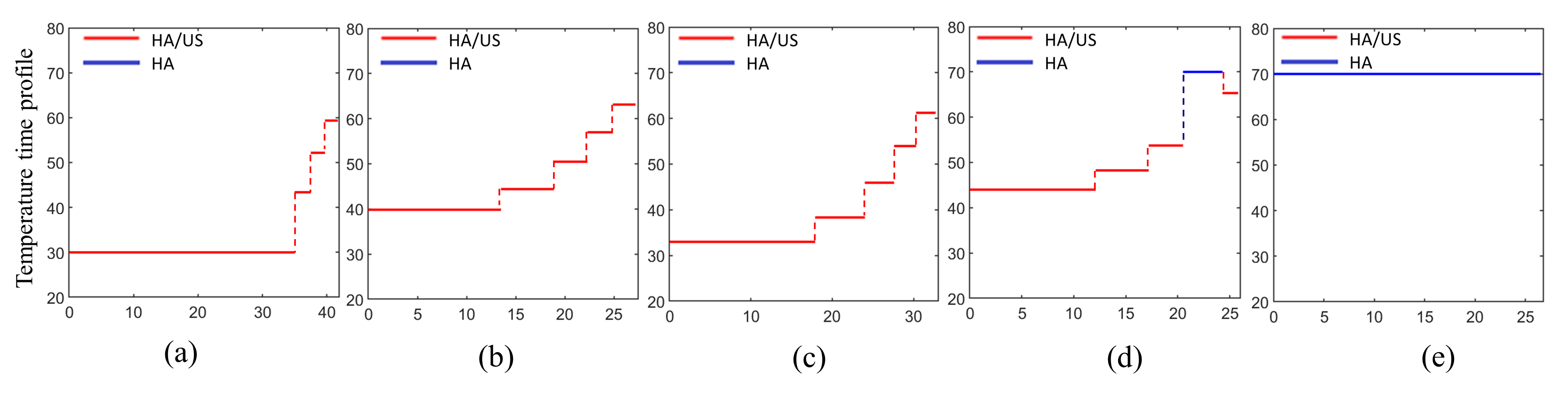}
  \caption{Figures (a), (b), (c), (d), and (e) represent the solution of Algorithm \ref{alg:Main} for $M=5$ and $\alpha = 0.2$, $\alpha = 0.1$, $\alpha = \frac{1}{15}$, $\alpha = 0.05$ and $\alpha = 0.04$, respectively. Here, $\alpha$ denotes the relative cost weight of the pure HA process.}
   \label{fig:gamma_vary}
\end{figure*}

In both steps, the value of $\Bar{F}$ is reduced. Therefore, the algorithm converges. Furthermore, we can adjust the relative weight of the entropy term $-H$ to the average cost $\Bar{D}$ using the Lagrange multiplier $\beta$. For $\beta \rightarrow 0$, maximizing the entropy term dominates minimizing the expected cost. In this case, the optimal weights derived in (\ref{eq:p_opt}) would be equal for all the valid process configurations. On the other hand, when $\beta \rightarrow \infty$, more importance is given to $\Bar{D}$. In other words, for very large values of $\beta$, we have:
\begin{align}
\begin{split} \label{eq:p_lim}
\lim_{\beta \rightarrow \infty}p(\omega) = \begin{cases}
1 \quad \text{if} ~\omega = \underset{\omega^\prime \in \Omega}{\mathrm{\text{argmin}}}
 &D(\omega^\prime, \eta_1, \hdots, \eta_M)\\
 0 \quad \text{otherwise}
\end{cases}
\centering
\end{split}
\end{align}
The idea behind the algorithm is to start with $\beta$ values close to zero, where the objective function $\Bar{F}$ is convex and the global minimum can be found. Then, we keep track of this global minimum by gradually increasing $\beta$ until $\max_{\omega \in \Omega} p(\omega) \rightarrow 1$. This procedure helps us avoid poor local minima. The proposed algorithm is shown in Algorithm \ref{alg:Main}.
\begin{algorithm}[t]
\caption{Combinatorial Optimization using MEP}
\begin{algorithmic} \label{alg:Main}
\STATE \textbf{Initialize:} $\eta_k \in U$, $\beta = \beta_{min}$, $\epsilon \approx 1^-$, $\zeta \geq 1$
\STATE Compute $p(\omega)$ for all $\omega \in \Omega$ using (\ref{eq:p_opt}).
\WHILE{$\max_{\omega}p(\omega) \leq 1 - \epsilon$}
\STATE Find optimal $\{\eta_k^*\}$ using any constrained optimization algorithm to solve (\ref{eq:eta_opt}).
\STATE Update $\{\eta_k\} \leftarrow \{\eta_k^*\} \quad \forall 1 \leq k \leq M$.
\STATE Update $p(\omega)$ using (\ref{eq:p_opt}) for all $\omega \in \Omega$.
\STATE Set $\beta \leftarrow \zeta \beta$.
\ENDWHILE
\end{algorithmic}
\end{algorithm}

\section{Simulations and Results}\label{sec:SR}

In this section, we simulate our proposed algorithm for drying DDG products using multiple sub-processes and compare it to the commonly-used single-stage drying process. By changing the number of sub-processes allowed ($M$), we investigate how additional sub-processes affect efficiency. Moreover, we can also assign weights to the energy consumed by different sub-processes to include their additional costs (e.g., maintenance), using the coefficient $\alpha$ defined in (\ref{eq:Cost_HA}) in our problem formulation.

\begin{wrapfigure}{r}{0.23\textwidth}
  \centering
  \includegraphics[width=0.23\textwidth]{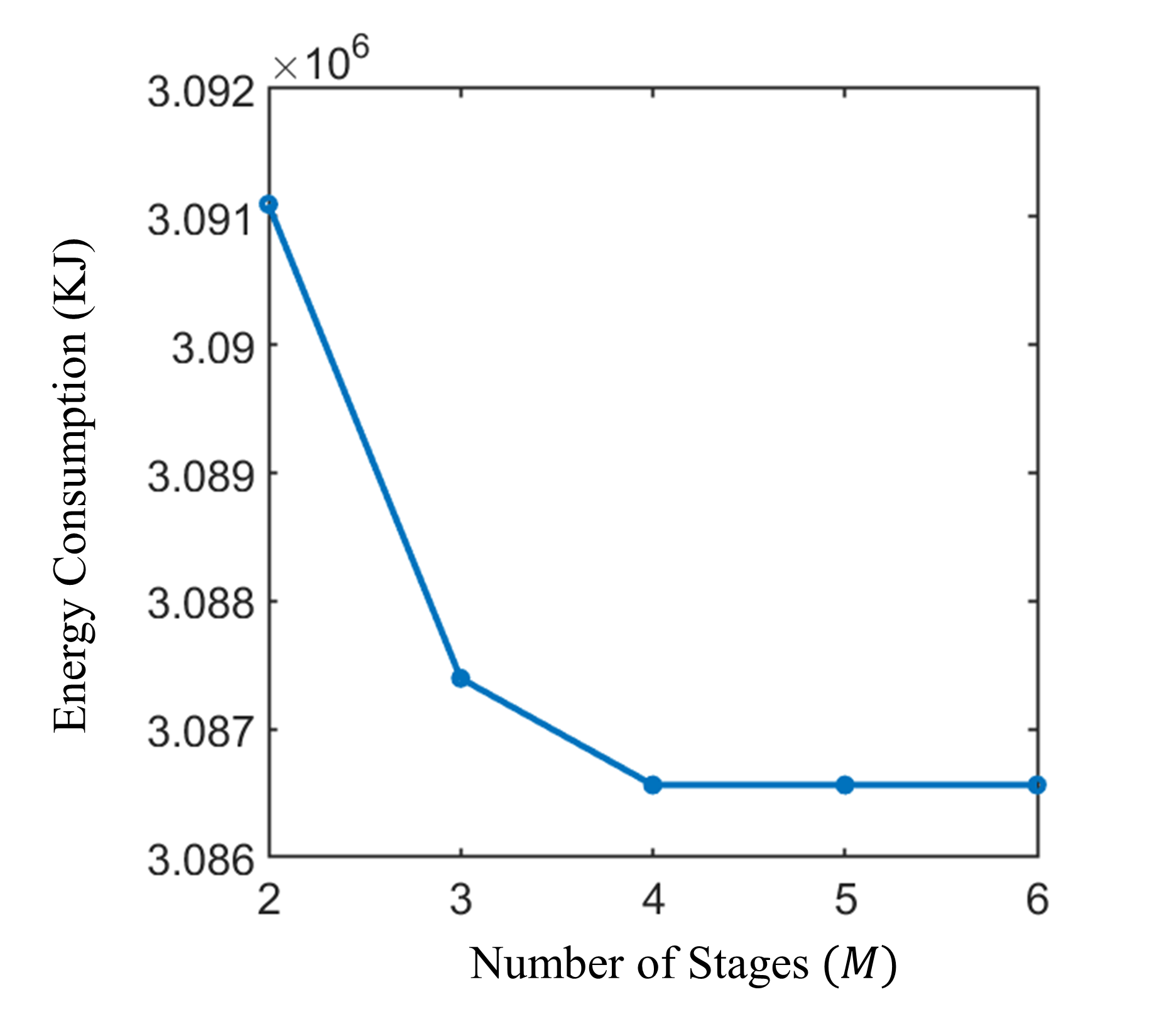}
  \caption{Energy consumption of the algorithm solution for different numbers of allowable stages.}
  \label{fig:Er_gamma0.5}
\end{wrapfigure}

\textbf {Effect of the permissible number of stages ($M$):} In simulations shown in Fig. \ref{fig:gamma0.5}, we have considered drying fresh DDG products with roughly $75.5\%$ initial wet basis moisture content to around $7.5\%$ at the end of the process. We have used our proposed algorithm for $\alpha = 0.5$ with different numbers of allowable sub-processes from two to six (Fig. \ref{fig:Er_gamma0.5}).
The results show 11.96\% ($M=2$), 12.07\% ($M=3$), and 12.09\% ( $M=4$, $M=5$, and $M=6$) improvement in energy consumption compared to the most efficient single-stage HA/US drying process. In addition, it reduced the energy consumption by  63.13\%  ($M=2$), 63.18\% ($M=3$), and 63.19\% ($M=4$, $M=5$, and $M=6$), in comparison with the optimal pure HA process.

As shown in Fig. \ref{fig:Er_gamma0.5}, the cost of the solution given by Algorithm \ref{alg:Main} decreases as the number of allowable sub-processes increases from two to four. However, for $M \geq 4$, increasing $M$ does not impact energy consumption reduction. 
In general, by increasing $M$, the cost of the process either decreases or remains constant. The reason is that the space of all valid process configurations with $M$ stages is a subspace of process configurations with $N > M$ stages. Consequently, we can choose $M$ in such a way that further increasing it does not significantly affect the cost value. In this case, as an example, $M=4$ can be chosen as the optimal number of stages.

\textbf{Effect of the relative weight of HA process ($\alpha$):} With $\alpha = 0.5$, the HA/US process is significantly more efficient than the pure HA process, resulting in the latter being absent from the optimal solutions in Fig. \ref{fig:gamma0.5}. As $\alpha$ is reduced, Fig. \ref{fig:gamma_vary} shows that the pure HA process becomes a part of the optimal solution, with optimal configurations consisting only of HA/US sub-processes for $\alpha = 0.2$, $\alpha = 0.1$, and $\alpha = \frac{1}{15}$. For $\alpha = 0.05$, the pure HA sub-process is used in one stage, and for $\alpha = 0.04$, the optimal solution is entirely a pure HA process with a constant temperature profile. The solutions (a)-(d) in Fig. \ref{fig:gamma_vary} achieve energy consumption reductions of 9.95\%, 5.75\%, 3.62\%, and 2.08\% compared to the most efficient single-stage processes.

\section{Conclusion and Future Work}

In this paper, we introduced a class of combinatorial optimization problems prevalent in industrial processes involving sub-processes with similar objectives. We focused on industrial drying, examining continuous and batch processes, and applied our proposed algorithm to a batch process drying prototype that allowed for both HA and HA/US drying. Our study demonstrated the benefits of simultaneous optimization of the process configuration and control parameters, as opposed to treating them as separate problems.

Although our example was limited to two permitted technologies, our framework can be extended to accommodate any number of technologies $|\Gamma_k| \in \mathbb{N}$ for all $1 \leq k \leq M$. Additionally, our algorithm can be modified to include more control parameters and quality constraints. In future work, we plan to include air velocity, ultrasound power, and duty cycle as control variables, and quantitative color as a constraint representing desired features.

The methodology we presented yielded a combinatorially large space of decision variables, with a complexity of $O(\sum_{k=1}^{M} {N \choose k})$. To reduce this complexity, we plan to employ the Principle of Optimality in our future work. This principle states that the next technology and its operating conditions are determined only by the current state, independent of prior sub-processes, along an optimal sequence of sub-processes. Successfully utilizing this fact will increase the scalability of our proposed algorithm. Additionally, the algorithm can be adjusted to incorporate new constraints specific to the technologies and setup used.

\bibliographystyle{IEEEtran}
\bibliography{IEEEabrv}

\begin{thebibliography}{10}
\providecommand{\url}[1]{#1}
\csname url@rmstyle\endcsname
\providecommand{\newblock}{\relax}
\providecommand{\bibinfo}[2]{#2}
\providecommand\BIBentrySTDinterwordspacing{\spaceskip=0pt\relax}
\providecommand\BIBentryALTinterwordstretchfactor{4}
\providecommand\BIBentryALTinterwordspacing{\spaceskip=\fontdimen2\font plus
\BIBentryALTinterwordstretchfactor\fontdimen3\font minus
  \fontdimen4\font\relax}
\providecommand\BIBforeignlanguage[2]{{%
\expandafter\ifx\csname l@#1\endcsname\relax
\typeout{** WARNING: IEEEtran.bst: No hyphenation pattern has been}%
\typeout{** loaded for the language `#1'. Using the pattern for}%
\typeout{** the default language instead.}%
\else
\language=\csname l@#1\endcsname
\fi
#2}}

\bibitem{DOEBarrier}
``Barriers to industrial energy efficiency,'' US Department of Energy, Tech.
  Rep., June 2015.

\bibitem{DOEReview}
``Quadrennial technology review,'' US Department of Energy, Tech. Rep.,
  September 2015.

\bibitem{NutritionalP}
N.~nan An, W.~hong Sun, B.~zheng Li, Y.~Wang, N.~Shang, W.~qiao Lv, D.~Li, and
  L.~jun Wang, ``Effect of different drying techniques on drying kinetics,
  nutritional components, antioxidant capacity, physical properties and
  microstructure of edamame,'' \emph{Food Chemistry}, vol. 373, p. 131412,
  2022.

\bibitem{DEP2}
M.~Yang and J.~Yagoobi, ``Enhancement of drying rate of moist porous media with
  dielectrophoresis mechanism,'' \emph{Drying Technology}, vol.~0, no.~0, pp.
  1--12, 2021.

\bibitem{US}
A.~Malvandi, D.~{Nicole Coleman}, J.~J. Loor, and H.~Feng, ``A novel
  sub-pilot-scale direct-contact ultrasonic dehydration technology for
  sustainable production of distillers dried grains (\uppercase{DDG}),''
  \emph{Ultrasonics Sonochemistry}, vol.~85, p. 105982, 2022.

\bibitem{US2}
O.~Kahraman, A.~Malvandi, L.~Vargas, and H.~Feng, ``Drying characteristics and
  quality attributes of apple slices dried by a non-thermal ultrasonic contact
  drying method,'' \emph{Ultrasonics Sonochemistry}, vol.~73, p. 105510, 2021.

\bibitem{SJR}
M.~Farzad and J.~Yagoobi, ``Drying of moist cookie doughs with innovative slot
  jet reattachment nozzle,'' \emph{Drying Technology}, vol.~39, no.~2, pp.
  268--278, 2021.

\bibitem{IR}
D.~Huang, P.~Yang, X.~Tang, L.~Luo, and B.~Sunden, ``Application of infrared
  radiation in the drying of food products,'' \emph{Trends in Food Science \&
  Technology}, vol. 110, pp. 765--777, 2021.

\bibitem{DEP}
\emph{{Experimental Study of Heat Transfer Characteristics of Drying Process
  with Dielectrophoresis Mechanism}}, ser. ASME International Mechanical
  Engineering Congress and Exposition, vol. Volume 11: Heat Transfer and
  Thermal Engineering, 11 2021.

\bibitem{RSM2}
Z.~Erbay and F.~Icier, ``Optimization of hot air drying of olive leaves using
  response surface methodology,'' \emph{Journal of Food Engineering}, vol.~91,
  no.~4, pp. 533--541, 2009.

\bibitem{RSM3}
H.~Majdi, J.~Esfahani, and M.~Mohebbi, ``Optimization of convective drying by
  response surface methodology,'' \emph{Computers and Electronics in
  Agriculture}, vol. 156, pp. 574--584, 2019.

\bibitem{Kale}
N.~V. Kale and S.~M. Salapaka, ``Maximum entropy principle-based algorithm for
  simultaneous resource location and multihop routing in multiagent networks,''
  \emph{IEEE Transactions on Mobile Computing}, vol.~11, no.~4, pp. 591--602,
  2012.

\bibitem{Lewis}
W.~K. Lewis, ``The rate of drying of solid materials.'' \emph{Journal of
  Industrial \& Engineering Chemistry}, vol.~13, no.~5, pp. 427--432, 1921.

\bibitem{9096570}
A.~Srivastava and S.~M. Salapaka, ``Simultaneous facility location and path
  optimization in static and dynamic networks,'' \emph{IEEE Transactions on
  Control of Network Systems}, vol.~7, no.~4, pp. 1700--1711, 2020.

\bibitem{jaynes1957information}
E.~T. Jaynes, ``Information theory and statistical mechanics,'' \emph{Physical
  review}, vol. 106, no.~4, p. 620, 1957.

\bibitem{jaynes2003probability}
------, \emph{Probability theory: The logic of science}.\hskip 1em plus 0.5em
  minus 0.4em\relax Cambridge university press, 2003.

\end{thebibliography}

\end{document}